\documentclass[%
 reprint,
 superscriptaddress,
nofootinbib,
 amsmath,amssymb, amsfonts,
 showkeys,
prl
]{revtex4-1}

\usepackage{graphicx}% Include figure files
\usepackage{dcolumn}% Align table columns on decimal point
\usepackage{bm}% bold math
\usepackage{physics}
\usepackage[dvipsnames]{xcolor}
\usepackage[
    colorlinks,
    citecolor=RoyalBlue,
    linkcolor=RoyalBlue,
    urlcolor=RoyalBlue,
]{hyperref}
\usepackage{multirow}
\usepackage{mathtools}
\usepackage{verbatim}  % for comments
\usepackage[normalem]{ulem}

\begin{document}

% \title{Mie voids and Babinet’s principle in dielectric nanoresonators}
\title{Quasi-Babinet principle in dielectric resonators and Mie voids}

\author{Masoud Hamidi}
\thanks{Authors contributed equally}
\affiliation{Institute of Physics, University of Graz, and NAWI Graz, Universitätsplatz 5, Graz 8010, Austria}

\author{Kirill Koshelev}
\thanks{Authors contributed equally}
\email{kirill.koshelev@anu.edu.au}
\affiliation{Nonlinear Physics Center, Research School of Physics, Australian National University, Canberra, ACT 2601, Australia}

\author{Sergei Gladyshev}
\affiliation{Institute of Physics, University of Graz, and NAWI Graz, Universitätsplatz 5, Graz 8010, Austria}

\author{Adrià Canós Valero}
\affiliation{Institute of Physics, University of Graz, and NAWI Graz, Universitätsplatz 5, Graz 8010, Austria}

\author{Mario~Hentschel}
\affiliation{4$^{th}$  Physics Institute and SCoPE, University of Stuttgart, Pfaffenwaldring 57, 70569, Stuttgart, Germany}

\author{Harald Giessen}
\affiliation{4$^{th}$ Physics Institute and SCoPE, University of Stuttgart, Pfaffenwaldring 57, 70569, Stuttgart, Germany}

\author{Yuri Kivshar}
\affiliation{Nonlinear Physics Center, Research School of Physics, Australian National University, Canberra, ACT 2601, Australia}

\author{Thomas Weiss}
\email{thomas.weiss@uni-graz.at}
\affiliation{Institute of Physics, University of Graz, and NAWI Graz, Universitätsplatz 5, Graz 8010, Austria}
\affiliation{4$^{th}$ Physics Institute and SCoPE, University of Stuttgart, Pfaffenwaldring 57, 70569, Stuttgart, Germany}

\date{\today} % Leave empty to omit a date

% \begin{abstract}
% Mie voids emerged recently as a platform for confining electromagnetic waves with wavelengths down to the ultraviolet regime in air.  Since Mie voids are considered as a counterpart of high-index dielectric Mie resonators, it is natural to wonder about validity of Babinet's principle, which has been developed for plasmonic systems and states that the light scattering from an opaque body can be reconstructed from that of a hole with the same size and shape. Here, we study Babinet’s conditions for Mie voids and high-index Mie resonators. Using numerical and analytical calculations, we find that Babinet's principle is applicable for dielectric systems within certain boundaries, which we demonstrate for spherical and more generically shaped Mie resonators. Limitations arise due to geometry-dependent terms and absorption losses in Mie resonators and Mie voids. Thus, our work establishes simple design rules for constructing dielectric resonators with complex functionalities from their complementary counterparts.
% \end{abstract}

\begin{abstract}
    Advancing resonant nanophotonics requires novel building blocks. Recently, cavities in high-index dielectrics have been shown to resonantly confine light inside a lower-index region. These so-called {\it Mie voids} represent a counterpart to solid high-index dielectric Mie resonators, offering novel functionality such as resonant behavior in the ultraviolet spectral region. However, the well-known and highly useful Babinet’s principle, which relates the scattering of solid and inverse structures, is not strictly applicable for this dielectric case as it is only valid for infinitesimally thin perfect electric conductors. Here, we show that Babinet’s principle can be generalized to dielectric systems within certain boundaries, which we refer to as {\it the quasi-Babinet principle} and demonstrate for spherical and more generically shaped Mie resonators. Limitations arise due to geometry-dependent terms as well as material frequency dispersion and losses. Thus, our work not only offers deeper physical insight into the working mechanism of these systems but also establishes simple design rules for constructing dielectric resonators with complex functionalities from their complementary counterparts.
\end{abstract}


\maketitle

\paragraph*{Introduction.} Babinet's principle is a fundamental concept of electromagnetism that establishes a direct relation between the scattering properties of infinitesimally thin perfect metallic conductors and complementary apertures in metal sheets~\cite{meixner1946babinetsche}. In optics, Babinet's principle can be reformulated for diffraction patterns produced by source radiation passing through complementary metallic screens with interchanged series of holes and obstructions~\cite{booker1946slot}, leading to an exchange between transmission and reflection for perfect conductors. The generalized form of Babinet's principle provides an approximate relation between transmission and reflection from flat absorbing scatterers and complementary apertures in absorbing media~\cite{neugebauer1957extension}.
 
In the early 2000s, Babinet's principle was applied for the design of plasmonic metamaterials and metasurfaces~\cite{falcone2004babinet, al2008applying}. It was shown that the latter can be re-formulated for the field profiles of the electromagnetic resonances, connecting the electric and magnetic fields of normal and inverted structures~\cite{zentgraf2007babinet}. In particular, electric and magnetic fields exchange their roles inducing a flip between TE and TM polarizations~\cite{liu2008magnetoinductive}. The Babinet principle was also explored in the framework of the covariant coupled-dipole method~\cite{zhukovsky2013dichroism}. Later, the concept of self-complementary metasurfaces was proposed based on the Babinet principle and electromagnetic duality~\cite{nakata2013plane,ortiz2013self,urade2015frequency,baena2015self}. Based on that concept, a new type of surface waves in self-complementary metasurfaces was suggested and extensively studied~\cite{singh2009spiral,gonzalez2014surface,yermakov2021surface}. In recent years, the engineering of resonant metasurfaces based on Babinet's principle was actively employed in plasmonics~\cite{ortiz2021extension,ortiz2021babinet}. More specifically, it was applied for creating double negative index materials~\cite{zhang2013creating}, manipulating polarization~\cite{zalkovskij2013optically}, enabling topological properties~\cite{dia2017guiding,bisharat2019electromagnetic}, wavefront control~\cite{ni2013ultra}, filtering~\cite{urade2016dynamically}, coherent perfect absorption~\cite{urade2016broadband}, observing plasmonic electromagnetically induced transparency~\cite{liu2010planar,hentschel2013babinet}, magnetic near-field imaging~\cite{horak2019limits}, and simultaneously realizing magnetic and electric hotspots~\cite{hrtovn2020plasmonic}. 

Dielectric metaphotonics emerged recently as a promising alternative to plasmonics, featuring high-index dielectric and semiconductor nanoresonators as building blocks of photonic structures~\cite{won2019into,kivshar2022rise}. Resonant dielectric metastructures supporting geometric Mie resonances were shown to exhibit artificial magnetic response, leading to versatile interference effects and strong field confinement in the volume of material, useful for nonlinear and quantum applications~\cite{koshelev2020dielectric}. Mie resonances can be engineered in individual nanoresonators or their arrays in the form of metasurfaces and photonic crystal slabs. They manifest as localized modes or nonlocal lattice modes, and give rise to unusual optical phenomena, such as bound states in the continuum~\cite{hsu2016bound,koshelev2018asymmetric}, anapoles \cite{miroshnichenko2015nonradiating,canos2021theory}, or directional scattering \cite{liu2018generalized}. 

Very recently, the concept of Mie voids was proposed in dielectric metaphotonics ~\cite{hentschel2023dielectric}. Counter-intuitively, it was found that low-index materials inside a high-index environment, such as air voids in silicon, can support Mie resonances in the infrared, visible, and even ultraviolet range due to confinement of light inside the air region. It was also shown that such Mie void modes are robust to geometrical and environmental changes, so that these modes can be realized experimentally by bringing the voids to the surface of thick silicon wafers. The voids exhibit pronounced features in scattering, leading to the generation of bright naturalistic colors with a high resolution~\cite{hentschel2023dielectric}. In striking difference to photonic crystal slabs and membranes based on collective lattice resonances~\cite{ossiander2023extreme}, even single Mie voids provide highly local resonant properties operating as individual pixels. The existence of such modes suggests that there might be an extension of Babinet's principle to lossless dielectric materials, defying conventional wisdom. Should it be established, it could unlock new degrees of freedom for the design of dielectric metadevices at the nanoscale.

\begin{figure}[t]
\includegraphics[width=0.95\linewidth]{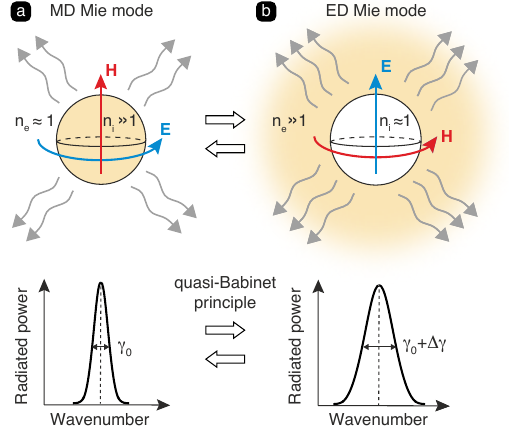}% Here is how to import EPS art
\caption{General concept. Quasi-Babinet principle establishes a correspondence between mode properties of (a) a high-index dielectric structure and (b) a low-index void in high-index dielectric host medium as its inverse system. The void modes exhibit a linewidth change $\Delta\gamma$ with respect to the mode of the normal structure linewidth $\gamma_0$, schematically shown with gray arrows (top panels) and the radiated power (bottom panels). The quasi-Babinet principle connects modes of orthogonal polarizations, such as a magnetic dipole (MD) and an electric dipole (ED) Mie mode with their electric and magnetic fields, $\mathbf{E}$ and $\mathbf{H}$ (schematically displayed by blue and red arrows), respectively.}
\label{fig:1}
\end{figure}

In this Letter,  we study similarities and differences of Mie modes in voids and high-index nanoresonators. We reveal that for spherical nanoparticles, a \emph{quasi-Babinet principle} can be established, connecting the electric and magnetic fields of a structure and its complementary counterpart, as well as their resonant frequencies and quality factors. The new principle holds quantitatively for voids (nanoparticles) with size larger than the wavelength, and approximately for subwavelength voids (nanoparticles). The principle remains approximately valid for non-spherical geometries, except in the vicinity of an avoided resonance crossing. Finally, we calculate the mode parameters for realistic dielectric materials with material losses in the visible and ultraviolet spectral range and demonstrate that the Babinet's principle is not applicable in the region of high absorption due to different loss rates for voids and nanoparticles, as intuitively expected.

\paragraph*{The quasi-Babinet principle for spherical resonators.} Babinet's principle offers a framework for correlating the scattering characteristics of two complementary structures. We emphasize that this principle is strictly valid only when both structures are fabricated from infinitesimally thin, metallic perfect conductors. In contrast, for dielectric materials with weak absorption used in dielectric nanophotonics, Babinet's principle is generally not applicable. We consider two complementary dielectric structures that are schematically shown in the top panel of Fig.~\ref{fig:1}: a high-index sphere with internal refractive index $n_\text{i}$ immersed in a low-index exterior composed of a medium with refractive index $n_\text{e}\ll n_\text{i}$ and a low-index void in a high-index host medium with $n_\text{e}\gg n_\text{i}$. We now study the applicability of Babinet's principle for these two scenarios. In the following, we use the term ``normal structures" referring to dielectric particles, and ``inverse structures" when referring to the voids embedded in a dielectric medium.

We start by investigating the behavior of the eigenmodes supported by the normal and inverse structures.  The latter can be classified into transverse electric (TE) and transverse magnetic (TM). Mie theory~\cite{Mie1908a, hentschel2023dielectric} yields the complex resonant wavenumbers $k_0=\omega/c$ for the normal and inverse cases, as solutions to the following transcendental equations:
\begin{align} 
	&\mathrm{TM\ modes\colon} & \frac{\psi_l'(n_\mathrm{i}k_0R)}{\psi_l(n_\mathrm{i}k_0R)}&=\frac{n_\mathrm{i}}{n_\mathrm{e}}\frac{\xi_l'(n_\mathrm{e}k_0R)}{\xi_l(n_\mathrm{e}k_0R)},\label{TM}\\ 
	&\mathrm{TE\ modes\colon} & \frac{\psi_l'(n_\mathrm{i}k_0R)}{\psi_l(n_\mathrm{i}k_0R)}&=\frac{n_\mathrm{e}}{n_\mathrm{i}}\frac{\xi_l'(n_\mathrm{e}k_0R)}{\xi_l(n_\mathrm{e}k_0R)}.\label{TE}
\end{align}
Here, $l=1,2,\dots$ is the orbital mode index, and $n_\mathrm{e}$ and $n_\mathrm{i}$ are the refractive indices outside and inside the sphere with radius $R$, respectively. The prime denotes derivatives with respect to the argument, and $\psi_l$ as well as $\xi_l$ are the Riccati-Bessel functions of order $l$, with $\psi_l(x)=xj_{l}(x)$ and $\xi_l(x)=xh_{l}(x)$, where $j_l$ and $h_l$ are the spherical Bessel and outgoing spherical Hankel functions, respectively. We note that the modes of spherical particles are degenerate with respect to the azimuthal index $m=0,\pm1,\pm2,\hdots,\pm l$.

We can expand the left and right part of Eqs.~(\ref{TM},\ref{TE}) in a series with respect to $1/z$ for $z=n_\text{i}k_0R$, assuming $z\gg1$, and solve the equation up to a specific order of this parameter. For normal ($n_\text{e}\ll n_\text{i}$) and inverse ($n_\text{e}\gg n_\text{i}$) structures, the solutions up to second order are (see Supplementary Material, Sec. I.A~\cite{SM})
\begin{align}
	z_{ls}\!&\approx\!  \tilde{z}_{l+1,s}\!-\!\frac{l(l\!+\!1)}{2\tilde{z}_{l+1,s}}\!\mp\!i\frac{n_\text{i}}{n_\text{e}}\frac{l(l\!+\!1)}{2\tilde{z}_{l+1,s}^2}& &\hspace{-0.3cm}\text{\ for }\hspace{-0.15cm}\begin{cases}\text{TM\ normal} \\ \text{TE\ inverse}\end{cases}\hspace{-0.4cm},
	\label{TM1norm}\\
	z_{ls}\!&\approx\! \tilde{z}_{ls}-\frac{l(l\!+\!1)}{2\tilde{z}_{ls}}\!\pm\!i\frac{n_\text{i}}{n_\text{e}}\frac{l(l\!+\!1)}{2\tilde{z}_{ls}^2}& &\hspace{-0.3cm}\text{\ for }\hspace{-0.15cm}\begin{cases}\text{TE\ normal} \\ \text{TM\ inverse}\end{cases}\hspace{-0.4cm}.\label{TE1norm}
\end{align}
Here $s=1,2,\dots$ is the radial mode index and $\tilde{z}_{ls}=\pi\left(s+(l-1)/2\right)-i\mathrm{arctanh}\left(n_</n_>\right)$ is the zeroth order solution for $n_<$ and $n_>$ as the lower and higher refractive indices, respectively. We note that the expansion in Eqs.~(\ref{TM1norm},\ref{TE1norm}) up to the zeroth order was known in the literature~\cite{sztranyovszky2022optical}, but without acknowledging the similarities between normal and inverse structures.

\begin{figure}
\includegraphics[width=0.95\linewidth]{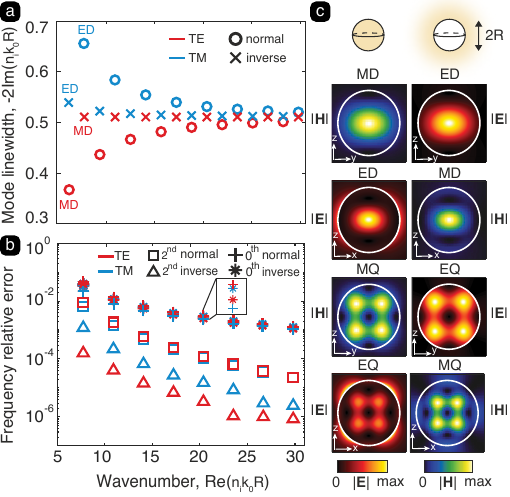}% Here is how to import EPS art
\caption{Demonstration of the quasi-Babinet principle for a high-index dielectric sphere with $n_{i}=4$, $n_e=1$ and a spherical low-index void with $n_i=1$, $n_e=4$. (a) Mode wavenumber $\operatorname{Re}{(n_ik_0R)}$ vs. mode linewidth $-2\operatorname{Im}{(n_ik_0R)}$ for $l=1$ obtained via solving Eqs.~(\ref{TM},\ref{TE}). (b) Relative error $|1-z_{ls}/z^{\rm ana}_{ls}|$ between the approximate solution $z_{ls}$ of Eqs.~(\ref{TM1norm},\ref{TE1norm}) for different orders of approximation and $z^{\rm ana}_{ls}$ as solutions of Eqs.~(\ref{TM},\ref{TE}) for $l=1$. (c)  Field profiles in normal and inverse structures for the fundamental MD and ED modes ($s=1$, $l=1$, $m=1$) labeled in (a) and fundamental magnetic quadrupolar (MQ) and electric quadrupolar (EQ) modes ($s=1$, $l=2$, $m=1$).}
\label{fig:2}
\end{figure}
%%%%%%
A comparison of TM results in the normal structure and TE results in the inverse structure reveals that the solutions are identical up to the first order of $1/z$. The same analogy can be made with TE results of the normal and TM results of the inverse structure. Therefore, we conclude that spherical dielectric nanoparticles and Mie voids are complementary up to the first-order in $1/z$. The difference in resonant wavenumbers for normal TE and inverse TM (or, normal TM and inverse TE) modes is given by the second-order correction, proportional to $l(l+1)/R^2$. We denote this approximate correspondence as the \textit{quasi-Babinet principle}. As a result of it, we will show that the modes of the complementary structures feature very similar resonant wavelengths, but different linewidths, as schematically depicted in the bottom panels of Figs.~\ref{fig:1}(a,b). 

In the Supplementary Material, we evaluate mode frequencies for a one-dimensional planar dielectric slab and air slot waveguide obtaining closed-form expressions, see~\cite[Sec.~I.B]{SM}. The analysis proves that the quasi-Babinet principle is exact for the slabs, since the mode frequencies have the same form as the zeroth-order approximation for the spheres above. 

Next, we evaluate the validity of the quasi-Babinet principle numerically by comparing solutions $z^{\rm ana}_{ls}$ of Eqs.~(\ref{TM},\ref{TE}) for normal ($n_{\rm i}=4, n_{\rm e}=1$) and inverse ($n_{\rm i}=1, n_{\rm e}=4$) cases, corresponding to a silicon sphere in air and an air void in silicon in the near-infrared range, respectively. For solving the transcendental Eqs.~(\ref{TM},\ref{TE}), we used a custom Python code. Figure~\ref{fig:2}(a) depicts the mode linewidth with respect to the real part of resonant wavenumbers for $l=1$ and $m=1$. We note the studied modes can be excited at plane-wave incidence. The difference between the real part of $n_{\rm i}k_0R$ for normal TE (TM) and inverse TM (TE) modes is small (note the different scale of axis) and decreases with the increase of the radial mode index $s$ (from left to right). The difference between the linewidths of complementary modes is larger, especially, for normal TM and inverse TE modes. However, it also decreases with the increase of the radial mode index. Therefore, the quasi-Babinet principle accuracy increases with the increase of $s$. 

To quantify the validity of the approximate solutions in Eqs.~(\ref{TM1norm},\ref{TE1norm}), we calculated the relative error $|1-z_{ls}/z^{\rm ana}_{ls}|$ between the analytical wavenumbers $z^{\rm ana}_{ls}$ and the approximate zeroth- and second-order solutions $z_{ls}$ for both the normal and inverse cases, shown in Fig.~\ref{fig:2}(b).  The inverse TE modes (red triangles) exhibits the highest accuracy for the second-order solution compared to the other cases. We note Fig.~\ref{fig:2}(b) confirms that the zeroth-order solution can be used to predict the location of the real part of the resonant wavenumbers.

Figure~\ref{fig:2}(c) displays the absolute value of the electric field $\boldsymbol{\mathrm{E}}$ and magnetic field $\boldsymbol{\mathrm{H}}$ for lowest-order multipolar modes of the normal and inverse structures normalized to their maximum values, which are magnetic dipolar (MD) and electric dipolar (ED) modes with $s=1$, $l=1$ and magnetic quadrupolar (MQ) and electric quadrupolar (EQ) modes with $s=1$, $l=2$. One can see a close resemblance between the fields of the normal structure modes and the corresponding modes of the void structure, confirming the validity of the quasi-Babinet principle.

\begin{figure*}
 \centering
\includegraphics[width=0.85\linewidth]{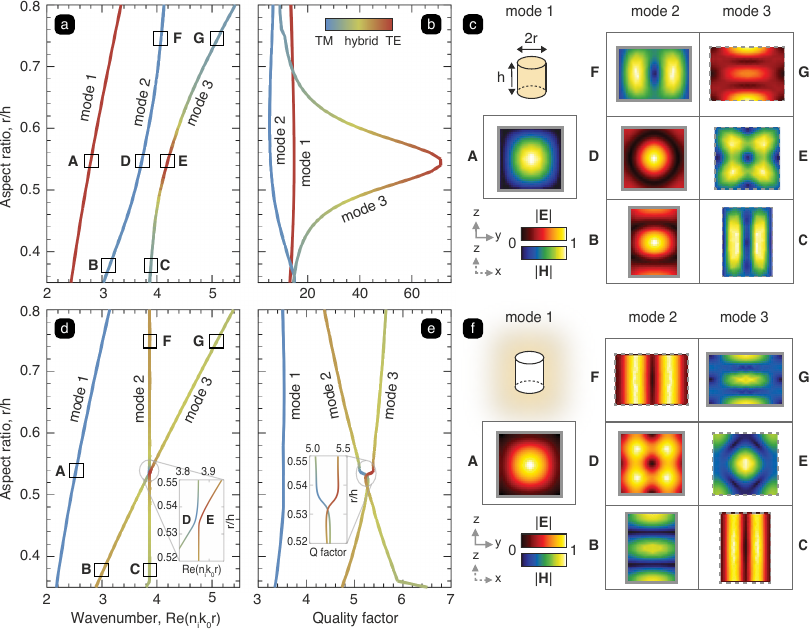}% Here is how to import EPS art
\caption{Quasi-Babinet principle for disk-shaped resonators: Mode wavenumber, quality factor, and field profiles vs. disk aspect ratio for the  fundamental modes of (a-c) normal ($n_{\rm i}=4$, $n_{\rm e}=1$) and (d-e) inverse structures ($n_{\rm i}=1$, $n_{\rm e}=4$), respectively. Only the eigenmodes that can be excited by a plane wave incident along the cylinder axis have been considered. The line color in (a,b,d,e) is defined by the mode's far-field polarization degree. The insets displays data within a narrow range of parameters near the avoided resonance crossing. The box line style in (c,f) denotes $zy$ (solid) and $zx$ (dashed) cross sections.}
\label{fig:3}
\end{figure*}

We would like to mention that the quasi-Babinet principle also holds for magneto-dielectric spherical structures with nonzero permeability and permittivity contrast  (see Supplementary Material~\cite[Sec.~II]{SM}). In this case, the modes of normal and inverse structures also obey the quasi-Babinet principle with the transition between ``normal" and ``inverse" defined via the inversion of the relative impedance between the internal and external domains of the sphere. 

% \sout{We also note that combining the electromagnetic duality considerations and the quasi-Babinet principle, one can establish a direct correspondence between the TE (TM) modes of normal dielectric structures and TE (TM) modes of mu-near-zero structures, and the same for TE (TM) modes of normal magnetic structures and TE (TM) modes of epsilon-near-zero structures, which may find specific applications in epsilon-and-mu-near-zero photonics~\cite{mahmoud2014wave}}.

\begin{figure}
\centering
\includegraphics[width=0.98\linewidth]{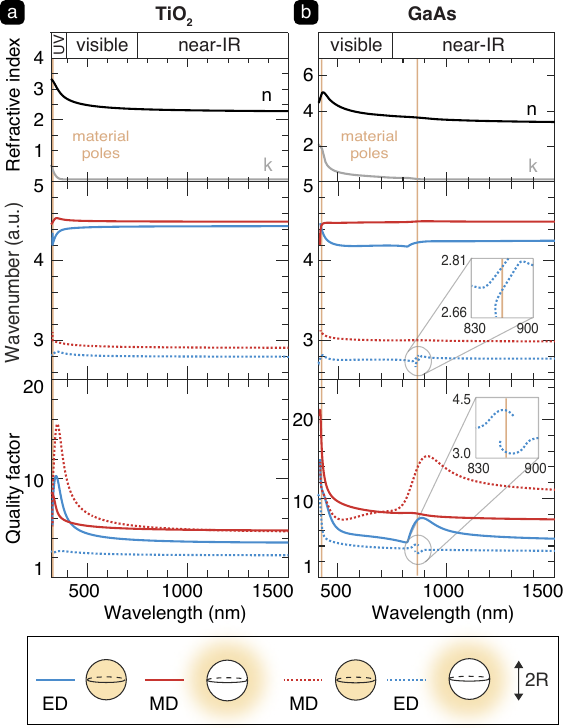}% Here is how to import EPS art
\caption{Effect of material absorption: Dependence of material refractive index, mode wavenumber $\operatorname{Re}{(n_{\rm i}k_0R)}$ and quality factor on resonant wavelength from the ultraviolet to near-infrared range for normal and inverse structures composed of air and (a) TiO$_2$, (b) GaAs. The bottom panel depicts the color legend for the ED and MD modes. The wavelengths of material poles are shown with brown solid lines.} 
\label{fig:4}
\end{figure}

\paragraph*{Non-spherical geometries.} We furthermore analyzed the applicability of the quasi-Babinet principle for resonators with non-spherical geometry. To do so, we compared the eigenmode spectrum of a dielectric disk with $n_\text{i}=4$ in air and a disk-shaped air void in an environment with $n_\text{e}=4$. Figure~\ref{fig:3} illustrates the mode wavenumber, quality factor, and field profiles of the first three fundamental modes of the normal and inverse disk structures as a function of the aspect ratio $r/h$ of radius $r$ and height $h$, respectively. For numerical calculations, we used the eigenmode solver in COMSOL Multiphysics\copyright. Unlike spheres, the eigenmodes of a cylinder have a mixed TE-TM character \cite{gladyshev2020symmetry,kuznetsov2022special}, i.e., they can radiate as mixtures of multipoles. We characterize their multipolar nature by performing a multipole decomposition of the eigenmode's radiated power into TE and TM contributions, which are then color-coded in Figs.~\ref{fig:3}(a,b,d,e) (see also Supplementary Material~\cite[Sec.~III]{SM}). 

The quasi-Babinet principle holds well for mode~$1$,   with a relative difference in wavenumber values below $13\%$ for the given range of $r/h$. Importantly, mode~$1$ exhibits a pure TE and TM nature for normal and inverse structures, respectively. The resonant frequencies of modes~$2$ and~$3$  avoid a crossing at $r/h\simeq0.535$, a clear signature of mode coupling. Despite these qualitative similarities, the quasi-Babinet principle no longer holds in the vicinity of the avoided crossing. 

In the normal structure, modal interference results in a decrease in the quality factor of mode~$2$, and a pronounced peak of the quality factor of mode~$3$, signaling the formation of a quasi bound state in the continuum~\cite{rybin2017high} of pure TE nature. This is in contrast to the inverse structure [Fig.\ref{fig:3}(d,e)]. On the one hand, the coupling between the modes is weaker, as can be seen from the much smaller separation between the dispersions of modes~$2$ and~$3$. On the other hand, the peak in the quality factor is less pronounced, but still mode~$3$ becomes a pure TE mode. We note that the observed (small) increase of quality factor for mode~$3$ at $r/h\simeq0.535$ is the first demonstration of the formation of quasi bound state in the continuum in individual void structures.
The field profiles for all modes at the avoided resonance crossing and away from it are displayed in Figs.~\ref{fig:3}(c,f). From the quality factor behavior and field profiles, we conclude that the quasi-Babinet principle still holds approximately for $r/h<0.5$ and $r/h>0.6$, i.e., when the hybrid modes are spectrally separated. 

\paragraph*{Effect of material losses.} We next analyze how the material losses affect the range of applicability of the quasi-Babinet principle. We calculate the resonant frequencies of fundamental modes for normal ($n_{\rm i}=n+ik$) and inverse ($n_{\rm e}=n+ik$) spherical structures composed of air and high-index materials with realistic dispersion, such as TiO$_2$ and GaAs. We use a custom Matlab code to solve Eqs.~(\ref{TM},\ref{TE}) with the material permittivity function extended to the complex frequency plane by fitting to several material poles (see Supplementary Material~\cite[Sec. XX]{SM}). Figure~\ref{fig:4} displays the material refractive index data, mode wavenumber $\operatorname{Re}{(n_{\rm i}k_0R)}$, quality factor for MD and ED modes of normal and inverse spherical resonators. Insets in Fig.~\ref{fig:4}(b) display a zoom into the spectral region, in which a material pole hybridizes with an optical resonance, resulting in two disconnected dispersion branches. 

Figure~\ref{fig:4} shows that the quasi-Babinet principle is valid for the low-loss range, and the complementary modes (normal ED with inverse MD, and normal MD with inverse ED) have very close wavenumbers. We see that in the range of high material losses as well as in the vicinity of material permittivity function poles, the quasi-Babinet correspondence breaks down. We also note that the void mode's quality factors are large in the high-loss regime, reaching values above $10$~\cite{hentschel2023dielectric}. Finally, we analyze how the absorption losses change the mode eigenfrequency for normal and inverse structures using the zeroth-order analytical expression in Eqs.~(\ref{TM1norm},\ref{TE1norm}) (see Supplementary Material~\cite[Sec.~IV]{SM}). We show that the difference in relative values of wavenumbers and quality factors for complementary normal and inverse structures in the low-loss or non-dispersive wavelength range is defined by $k/n$ and small. We also demonstrate that in the high-loss or dispersive range the relative difference can be drastically large, explained by the modes of inverse structures exhibiting absorption only in the narrow domain of surrounding media, compared to normal structure modes that are affected by the material losses in the whole particle volume.

\paragraph*{Conclusion.} We have established a quasi-Babinet principle that allows us to predict mode characteristics of dielectric Mie voids from dielectric Mie resonators and vise versa. We have determined analytically and numerically the applicability range of the quasi-Babinet principle depending on the spatial dimensionality, geometry, and material losses of the structure. We anticipate that extensions of this principle to magneto-dielectric structures may lead to further developments in epsilon-and-mu-near-zero photonics, as well as open new horizons for smart engineering of nanoscale metadevices comprising applications in strong light-matter interaction, biosensing, and quantum information processing.

\begin{acknowledgments}
K. Koshelev acknowledges Oleh Yermakov and Andrey Bogdanov for many useful comments and discussions. This work was supported by Bundesministerium f\"{u}r Bildung und Forschung, Deutsche Forschungsgemeinschaft, (SPP1839 “Tailored Disorder” and GRK2642 “Towards Graduate Experts in Photonic Quantum Technologies”), the Ministerium f\"{u}r Wissenschaft, Forschung und Kunst Baden-W\"{u}rttemberg (RiSC Project “Mie Voids”, ZAQuant), the Australian Research Council (grant DP210101292).

%M. Hamidi and K. Koshelev contributed equally to this %work. K.K. and T.W conceived the original idea. M.H., %K.K., A.C.V., S.G., and T.W. performed the simulations and modeling. All authors discussed, interpreted, and corroborated the results. All authors participated in the preparation and writing of the manuscript. K.K. and T.W supervised the work.
\end{acknowledgments}

\bibliography{Babinet}

\end{document}